\documentclass[preprint,authoryear,12pt]{elsarticle_mod}

\journal{Planetary and Space Science}

\usepackage{graphicx}
\usepackage{amsmath}
\usepackage{amssymb}

\usepackage{aas_macros} 

\begin{document}

\begin{frontmatter}


\title{Constraining the Cratering Chronology of Vesta}

\author{David P.~O'Brien}

\address{Planetary Science Institute, 1700 E.~Ft.~Lowell, Suite 106, Tucson, AZ 85719\\obrien@psi.edu}


\author{Simone Marchi}

\address{NASA Lunar Science Institute, Southwest Research Institute, Boulder, CO}

\author{Alessandro Morbidelli}

\address{Observatoire de la C\^ote d'Azur, CNRS, Nice, France}

\author{William F.~Bottke}

\address{NASA Lunar Science Institute, Southwest Research Institute, Boulder, CO}

\author{Paul M.~Schenk}

\address{Lunar and Planetary Institute, Houston, TX}

\author{Christopher T.~Russell}

\address{Institute of Geophysics and Planetary Physics, University of California, Los Angeles, CA}

\author{Carol A.~Raymond}

\address{Jet Propulsion Laboratory, California Institute of Technology, Pasadena, CA}
\address{\vspace*{0.4in}Accepted to Planetary and Space Science on May 1, 2014}
\address{1 Table and 7 Figures at end of Paper\vspace*{-0.2in}}


\begin{abstract}

Vesta has a complex cratering history, with ancient terrains as well as recent large impacts that have led to regional resurfacing.  Crater counts can help constrain the relative ages of different units on Vesta's surface, but converting those crater counts to absolute ages requires a chronology function.  We present a cratering chronology based on the best current models for the dynamical evolution of asteroid belt, and calibrate it to Vesta using the record of large craters on its surface.  While uncertainties remain, our chronology function is broadly consistent with an ancient surface of Vesta as well as other constraints such as the bombardment history of the rest of the inner Solar System and the Ar-Ar age distribution of howardite, eucrite and diogenite (HED) meteorites from Vesta.

\end{abstract}



\end{frontmatter}




\section{Introduction}
\label{sec_intro}

Even before NASA's Dawn Mission arrived at Vesta, it was known to have an interesting cratering history.  Hubble Space Telescope images revealed a large central-peak crater near its south pole estimated to be 460 km in diameter and 13 km deep, comparable in diameter to Vesta itself \citep{Thomas1997Sci}.  Vesta had long been suggested as the parent body of the HED (howardite, eucrite and diogenite) meteorites on the basis of spectral and geochemical evidence \citep{McCord1970Sci, Consolmagno1977GCA}.  Its location in the inner asteroid belt between the 3:1 mean-motion resonance with Jupiter and the $\nu_6$ secular resonance with Saturn is favorable for delivering material to near-Earth space \citep{Wisdom1985Nat, Migliorini1997MAPS}, and there is a dynamical family of related bodies (often called `vestoids') in the vicinity of Vesta, stretching toward those resonances \citep{Binzel1993Sci}.  The detection of the large south pole crater quite literally provided the `smoking gun' evidence linking Vesta to the vestoids and HEDs.  Studies of the HED meteorites showed that Vesta formed and differentiated early in Solar system history \citep[see][for a review]{McSween2011SSRv}, such that its surface should provide a record of some of the earliest times in Solar System history.  Thus, there were many reasons for the selection of Vesta as one of the targets of the Dawn Mission \citep{Russell2011SSRv}.

More detailed observations once Dawn arrived at Vesta revealed that there were in fact two overlapping large impact basins in the southern hemisphere, the larger and younger one known from the HST imaging now named Rheasilvia, and the older one named Veneneia \citep{Schenk2012Sci}.  Vesta showed a strong dichotomy between north and south, with the north being heavily cratered and the south showing relatively few craters \citep{Marchi2012Sci}.  Initial crater counts of the Rheasilvia and Veneneia basins placed their ages at approximately 1 Ga for Rheasilvia, and at least 2 Ga for Veneneia, with the uncertainty in the latter being due to the fact that it was somewhat disrupted during the formation of Rheasilvia \citep{Marchi2012Sci, Schenk2012Sci}.  The formation of these two large basins erased nearly all pre-existing craters in the southern hemisphere, but apparently left the northern hemisphere relatively undisturbed.

Rheasilvia has sharply-defined features such as ridged terrain on the crater floor and a prominent rim scarp \citep{Schenk2012Sci}, suggestive of a relatively young age.  Additional evidence for the relatively recent formation of Rheasilvia comes from the Vesta asteroid family, the `vestoids,' that are dynamically related to Vesta and were likely ejected in a large cratering event.  The size distribution of these bodies is quite steep compared to the background population \citep[eg.][]{Cellino1991MNRAS, Tanga1999Icar, Nesvorny2008Icar}, and \citet{Marzari1996AA, Marzari1999Icar} showed that the family would have to be less than about 1 Ga old, otherwise its size distribution would collisionally grind down and not remain as steep as observed.  Finally, the Gamma Ray and Neutron Detector (GRaND) on Dawn detected hydrogen on the surface of Vesta, which is interpreted to be exogenic, and found a much lower abundance of hydrogen within the perimeter of the Rheasilvia basin compared to the rest of the surface, suggesting that Rheasilvia formed and reset the surface relatively recently \citep{Prettyman2012Sci}.

Broadly, the sum of this evidence points to an ancient surface that has been modified over time by craters, including several large and relatively recent ones, such that Vesta's surface units span a wide range of ages.  While relative crater densities can be used to place these different units in a stratigraphic sequence, one would always like to be able to determine absolute ages.  In practice, however, this is not always a straightforward process.  \citet{Marchi2012Sci} and \citet{Schenk2012Sci} used estimates of the impact rate in the current main belt to estimate the ages of Rheasilvia and Veneneia, which can work reasonably well for surfaces dating back to about 3 Ga or so, as the impact rate in the main belt was likely fairly constant over that time period \citep[see][for a recent review of main-belt dynamical and collisional history]{OBrien2011SSRv}.  Prior to that, however, the impact rate was likely higher, such that there is no longer a linear relationship between age and crater density (this relation is generally termed a `chronology function' or `chronology curve').  Furthermore, as we discuss in Sec.~\ref{sec_bg}, the degree to which the impact rate has changed over time has been a matter of debate.

Here we present a chronology based on the current understanding of main-belt dynamical history, which we then constrain using the record of the largest impacts on Vesta.  We first give a brief review of the current understanding of the impact history of the inner Solar System in Sec.~\ref{sec_bg}.  In Sec.~\ref{sec_lun} we present a mathematical description of the lunar chronology as well as the ``lunar-like'' chronology proposed for Vesta by \citet{Schmedemann2014PSS}, and in Sec.~\ref{sec_m} we give a derivation of our model chronology for Vesta based on main-belt dynamics.  We present these curves in a normalized form, such that they can be used with different estimates of the crater production function.  Our estimate of the crater production function is then discussed in Sec.~\ref{sec_prodfunc}.  In Sec.~\ref{sec_apply} we calibrate the model chronology curve to Vesta, using measurements of the large crater population on its surface as a constraint.  We summarize the results and discuss the implications of this work in Sec.~\ref{sec_discuss}.

\section{Background on the Impact History of the Inner Solar System}
\label{sec_bg}

The lunar cratering record is the most well-studied in the Solar System, and benefits from the availability of radiometrically dated samples from some surfaces that have been studied with crater counts.  Despite this `ground truth', however, there remains some ambiguity.  Radiometric dating of samples of lunar impact melts \citep[eg.][]{Papanastassiou1971aEPSL, Papanastassiou1971bEPSL, Wasserburg1971EPSL, Turner1973Nat} showed an unexpected clustering of ages around 3.9 Ga and absence of earlier ages, which ran counter to the prevailing idea that the impacts on the moon and terrestrial planets were due the dynamically-decaying remnants of planet formation.  \citet{Tera1974EPSL} coined the term `terminal Lunar cataclysm' for this spike in impact activity, occurring $\sim$600 Ma after the formation of the Moon (it is also widely referred to as the Late Heavy Bombardment, or LHB).  Later studies of melts in lunar meteorites \citep[eg.][]{Cohen2000Sci} showed a similar clustering of impact ages around 3.9 Ga, although the distribution was more broad than that inferred from the Apollo samples.

\citet{Hartmann1975aIcar} suggested that the apparent lack of impact melt ages prior to 3.9 Ga may be due to what he termed the `stone-wall', in which there was a smoothly declining cratering rate, but prior to $\sim$3.9 Ga it was so intense that no surface rocks were able to escape resetting.  While this cannot be strictly true, given that many older lunar rocks exist that did not experience any impact melting, more complicated processes might have occurred in which, for example, there is a selection effect towards older impact melt rocks being more deeply buried and younger ones more likely to be found on the surface \citep{Hartmann2003MAPS}.  Further modeling is necessary to fully understand regolith evolution and the types of selection effects it can introduce.

Combining crater counts with known radiometric ages can shed some light on the problem, although it is dependent to some degree on how well we can measure the crater production populations and how well we actually know the age of the surfaces.  For example, the age of the Nectaris basin, a key marker in the lunar chronology system, is placed at $\sim$4.1 Ga by some groups and $\sim$3.9 Ga by others, depending on how the distribution of ages in Apollo 16 samples is interpreted.  Assuming that the age of Nectaris is represented by the 4.1 Ga age signature in the Apollo 16 samples, \citet{Neukum1982LPSC}, \citep[also][]{Neukum1994Hazards, Neukum2001SSRv} proposed that the impact rate was relatively constant back to $\sim$3.5 Ga, and exponentially increased back to the formation of the moon at 4.5 Ga (however, as that analysis is only based on samples dating back to 4.1 Ga, it does not provide a solid constraint on what happened before 4.1 Ga).  On the other hand, if the age of Nectaris is assumed to be represented by the 3.9 Ga signature in the Apollo 16 samples, the exponential decay would be much steeper and cannot be extrapolated back to 4.5 Ga without the moon accreting more than a lunar mass of material \citep{Ryder1990EOS, Stoffler2001SSRv, Ryder2002JGR}.  This argues that the decay could not have been monotonic, and some sort of impact cataclysm must have occurred.

The latest work suggests that none of the available samples actually originate from Nectaris, making direct dating of that basin impossible \citep{Norman2010GCA}.  \citeauthor{Norman2010GCA} find that the 4.1-4.2 Ga ages seen in the Apollo 16 samples may actually be from the Serenitatis basin, with Serenitatis ejecta thrown to the Apollo 16 site by the Imbrium impact.  The Serenitatis basin may be only slightly older than Nectaris \citep[eg.][]{Fassett2012JGR}, however, so the work of \citeauthor{Norman2010GCA} is consistent with a $\sim$4.1 Ga age for Nectaris.

While it's fair to say that the evidence for or against the declining flux and impact cataclysm models from cratering records and dated lunar samples is still not conclusive \cite[see, eg.][for a discussion]{Hartmann2000OEM, Chapman2007Icar}, other lines of evidence can provide additional constraints.  For example, \citet{Morbidelli2012EPSL} discuss the constraints from highly siderophile element abundances in lunar rocks \cite[eg.][]{Walker2004EPSL, Day2007Sci, Day2010EPSL}.  They use those constraints to infer a total impacting mass of $\sim 3.5 \times 10^{19}$ kg on the Moon since its formation, whereas the total impacting mass implied by extending the \citet{Neukum1982LPSC}, \citet{Neukum1994Hazards} and \citet{Neukum2001SSRv} bombardment history back to 4.5 Ga is roughly four times larger.  \citet{Bottke2007Icar} modeled the dynamical and collisional evolution of possible long-lived dynamical reservoirs in the inner Solar System to determine if a decaying population of planetesimals could remain massive enough and decay slowly enough to match the number of basins that formed on the moon between $\sim$3.8 and 4.1 Ga.  They found that, especially when collisional grinding was considered, there was no population capable of surviving long enough to explain the formation of those basins.  Both of these studies point to the need for a delayed bombardment, regardless of whether the actual age of Nectaris is 3.9 or 4.1 Ga.

Various mechanisms have been proposed to explain how a delayed increase in the impact flux could occur, such as the late formation of Neptune \citep{Levison2001aIcar}, the destabilization of a fifth terrestrial planet that originally existed between Mars and the asteroid belt \citep{Chambers2007Icar}, or the breakup of a Vesta-sized body in the Mars-crossing population \citep{Cuk2012Icar}.  The leading scenario is generally called the \textit{Nice Model} as all of the authors were working in Nice, France when it was developed \citep{Tsiganis2005Nat, Gomes2005Nat, Morbidelli2005Nat}.  In this model, Jupiter and Saturn formed interior to a mutual mean-motion resonance (MMR, initially proposed to be the 2:1 resonance but more likely even closer together inside their 3:2 resonance).  The scattering of remnant planetesimals in a disk beyond the giant planets caused Jupiter and Saturn to slowly diverge until they crossed the resonance, triggering an instability that among other things destabilized the asteroid belt, cleared out most of the remnant planetesimals, and caused the giant planets to rapidly migrate to their current orbital configuration.

In the original simulations, \citet{Gomes2005Nat} found that the migration of Jupiter and Saturn across their mutual 2:1 MMR would cause resonances to sweep through the asteroid belt, sending those asteroids onto potentially Earth-crossing orbits and depleting the mass of the belt by a factor of 10-20.  The impact rate on the Moon would spike fairly sharply, consistent with the lunar cataclysm scenario suggested by \citet{Tera1974EPSL}.  However, more detailed work with a wider range of simulations found that the best match to a range of Solar System constraints was obtained when Jupiter has an encounter with one of the ice giants (Uranus or Neptune) immediately following the resonance crossing, causing a much more rapid migration than would be obtained simply by planetesimal scattering \citep{Brasser2009AA, Morbidelli2009AA, Morbidelli2010AJ}.  This was termed the `Jumping Jupiter' scenario.  \citet{Minton2009Nat} also found that a very short timescale of Jupiter's migration was necessary in order to match constraints from the asteroid belt, consistent with that scenario.

\citet{Morbidelli2010AJ} showed that unlike in the original simulations of \citet{Gomes2005Nat} where the asteroid belt was depleted by a factor of 10-20, the depletion factor in the jumping Jupiter case was only about 2.  \citet{Minton2010Icar} showed that there may be an additional factor of $\sim$2 depletion due to the longer term chaotic diffusion of unstable asteroids following Jupiter's migration, but that still means that the mass of the asteroid belt prior to Jupiter's migration was only about 4x its current mass, and could not provide sufficient impactors to explain all of the large basins on the Moon, especially the oldest ones.

An extension of this scenario was proposed by \citet{Bottke2012Nat}, who suggested that the primordial asteroid belt extended inwards to roughly 1.8 AU (compared to roughly 2.1 AU now).  This innermost part of the primordial asteroid belt was termed the ``E-Belt,'' and would have been stable prior to the giant planet migration in the Nice Model, since the $\nu_6$ secular resonance that currently defines the inner edge of the asteroid belt would have been located beyond the asteroid belt when Jupiter and Saturn were closer together, interior to their mutual 3:2 MMR.  When the 3:2 MMR was crossed and the Nice Model instability occurred, the $\nu_6$ would have swept through the belt to its current location, making the E-Belt unstable.  Bodies from that population would be sent on Earth/Moon-crossing orbits and have a relatively high collision probability with the Earth and Moon, compared to bodies derived from further out in the asteroid belt.  Furthermore, the impacting population would experience a relatively slow decay with time, rather than the sharp spike in impacts produced in the early Nice Model simulations.  \citeauthor{Bottke2012Nat} found that the impacts from E-Belt bodies would create roughly 10 lunar basins between 3.7 and 4.1 Ga, broadly consistent with the number of basins in the \citet{Neukum1982LPSC}, \citet{Neukum1994Hazards} and \citet{Neukum2001SSRv} chronology since $\sim$4.1 Ga (with Nectaris likely being the first or one of the first basins formed by the destabilization of the E-Belt).

The E-Belt impacts would not be enough to explain all of the large basins on the moon, however, as several times as many basins formed prior to Nectaris than after it.  \citet{Morbidelli2012EPSL} showed that the best fit to the lunar cratering record is from a combination of a declining impact flux due to leftover planetesimals from terrestrial planet accretion, which would have formed the pre-Nectarian basins, combined with an increased flux starting around 4.1 Ga due to Jupiter's migration and the destabilization of the asteroid belt and E-Belt that would have formed the more recent basins.  \citet{Marchi2012EPSL} find that the crater size-frequency distribution (SFD) on and near the Nectaris basin is different from that on pre-Nectarian terrains, suggesting a transition in the impactor population around that time.  While there could be multiple interpretations of the nature of the transition, they show that it is consistent with a change in velocity of the impactors as the more highly-excited E-Belt population begins to dominate over the primordial impacting population around the time of Nectaris formation.

It is important to note that while there have been many historical disagreements about the interpretation of the lunar cratering record, this hybrid model of \citet{Morbidelli2012EPSL} is essentially consistent with the lunar chronology proposed by \citet{Neukum1982LPSC}, \citet{Neukum1994Hazards} and \citet{Neukum2001SSRv} back to $\sim$4.1 Ga (ie.~back to the earliest time for which we have samples that can potentially be associated with a specific areas of the lunar surface).  The primary uncertainties lie in what happened before that time.  Such a hybrid scenario was alluded to earlier by \citet{Hartmann2000OEM}, although never explicitly modeled.

\subsection{Implications for the Asteroid Belt}
\label{sec_bg_sub1}

It is common practice to scale the crater production rate from one planet to another in the inner Solar System using estimates of the orbital distribution of impactors (namely the Near-Earth Asteroids, NEAs) and scaling laws for crater production \citep[eg.][]{Ivanov2001SSRv}.  While this approach is not without its difficulties and uncertainties, it is generally reasonable given that the different planets are all targets being hit by a single source population, the NEAs.  The same approach can not necessarily be applied to scale the early lunar crater production rate to the asteroid belt, however.  The dynamical history that delivers the impactors from the main belt to the terrestrial planet region may imply a much different collisional history for bodies in the asteroid belt compared to the moon and terrestrial planets.  A simple example is that if the asteroid belt was suddenly reduced in mass by a factor of 2 at $\sim$4 Ga, with the asteroids being delivered to the terrestrial planet region, the impact rate in the asteroid belt would merely drop by a factor of 2, while bodies in the terrestrial planet region would experience a significant spike in their impact rate.

Another important factor is that the NEA size distribution may differ from that of the main belt, because the NEAs are derived from the main belt in part by the action of size-dependent forces, namely the Yarkovsky effect \citep[see, eg.][for a thorough review]{Bottke2006AREPS}.  In fact, \citet{Strom2005Sci} and \citet{Marchi2009AJ} find that there are two different crater SFDs on the Moon, one for older highland terrains, and a steeper SFD for younger maria terrains.  The implication is that impactors hitting the older terrains have a SFD that closely matches the main belt, and were derived from it by a size-independent process (such as the effects of Nice-Model resonance sweeping on the belt), while the younger surfaces are cratered primarily by NEAs, which have a steeper SFD than the main belt due to the action of the Yarkovsky effect.  Thus, the impactor SFD inferred from small craters on the moon, which are primarily counted on younger surfaces, is not necessarily the same as the impactor SFD in the asteroid belt itself.  In addition, NEAs are delivered primarily through resonances in the inner asteroid belt, but Vesta can be hit by central-belt and many outer-belt bodies as well, and all of those regions have somewhat different size distributions \citep[eg.][]{Jedicke1998Icar, Masiero2011ApJ}.

Instead, we must look at the dynamical history of the asteroid belt itself and use that to determine the impact rate vs.~time.  As described in Sec.~\ref{sec_bg_sub1}, the asteroid belt since $\sim$4.1 Ga likely experienced a factor of $\sim$4 depletion due to the combined effects of the resonance sweeping during the Nice Model instability \citep{Morbidelli2010AJ} and the subsequent decay of unstable asteroids \citep{Minton2010Icar}.  The E-Belt, while it dominated the impacts on the terrestrial planets, would have only been a relatively small fraction of the mass in the primordial belt.  Prior to $\sim$4.1 Ga, rather than increasing back in time, the mass of the asteroid belt may have remained relatively constant at $\sim$4 times the current mass going back several hundred million years.

At the earliest times, immediately following the formation of the Solar System, the asteroid belt may have had significantly more mass than it currently does \citep{Wetherill1992Icar, Petit2001Icar, OBrien2007Icar}, which would have been depleted over the first $\sim$100 Ma and led to an increased impact rate in the belt during that time.  Another possible contributor to the early impact rate in the asteroid belt could be leftover scattered planetesimals from the terrestrial planet region, although the effect of such bodies on the asteroid belt has not been fully quantified.  Regardless of the source of these earliest impactors, it is likely that there was a somewhat larger impact rate in the asteroid belt immediately following the formation of the Solar System, which would have decayed to $\sim$4 times the current rate and stayed at that level until $\sim$4.1 Ga, then decayed to its current rate following the destabilization of the asteroid belt and E-Belt.  This depletion at $\sim$4.1 Ga would correspond to the beginning of the Late Heavy Bombardment on the moon.

There are two additional sources of impactors that we do not consider here.  The giant planet migration and resulting depletion of the asteroid belt and E-belt around $\sim$4.1 is driven by the scattering a massive primordial trans-Neptunian disk of planetesimals, some of which will cross the inner Solar System \citep[eg][]{Gomes2005Nat, Levison2009Nat}.  However, there is not strong evidence for a significant cometary component to the impact flux in the inner Solar System. The SFD of ancient craters surfaces on the moon, for example, is consistent with being derived entirely from main belt impactors \citep{Strom2005Sci, Marchi2009AJ, Marchi2012EPSL}.  Because highly siderophile elements (HSEs, like the platinum-group elements) in the lunar crust would have been heavily depleted during its differentiation, current abundances of those elements are likely due to subsequent impacts.  Abundances of HSEs in the lunar crust show no signature of primitive, carbonaceous chondritic material like CI or CM chondrite (taken to be good proxies for cometary material), suggesting that comets did not play a major role in the lunar bombardment \citep[eg.][]{Kring2002JGR, Galenas2011LPSC}. Similar results are obtained from studies of actual projectile fragments in regolith breccias from the Apollo 16 site \citep{Joy2012Sci}.  The lack of evidence for cometary impactors could be due to disintegration of comets once they reach into the inner Solar System \citep[eg.][]{Sekanina1984Icar}, or perhaps the primordial trans-Neptunian disk was somewhat less massive  \citep[eg.][]{Nesvorny2013ApJ} than initially envisioned in \citet{Gomes2005Nat}.  \citet{Broz2013AA} model the formation and evolution of asteroid families under the influence of an LHB-era cometary bombardment, and find that comets could potentially lead to a significant impact flux, although to explain the number of observed families it is likely that 80\% of comets are disrupted due to close approaches to the Sun before they are able to impact the asteroids.  Hence, we ignore for now the possibility of cometary impactors on Vesta, although this should be revisited in the future pending better constraints on the impact flux.

It has also been proposed that the formation of Jupiter may have scattered planetesimals through the asteroid belt region and caused a `Jovian Early Bombardment' \citep{Turrini2011MNRAS, Turrini2014PSS}.  The actual bombardment rate in that model can be quite large, potentially intense enough to erode Vesta's surface, but varies significantly based on chosen initial conditions.  All of the impacts in that model would have occurred very early, and Vesta may have taken at least several million years to differentiate and form its crust \citep[see][for a review]{McSween2011SSRv}, so it is not clear that those impacts would actually be recorded on its surface as seen today.  Hence we do not include it in the model presented here, but note that scattered planetesimals during Jupiter's formation may be an additional impactor population to include in future analysis, provided it can be better constrained.

\section{The Lunar and ``Lunar-Like'' Chronologies}
\label{sec_lun}

The cumulative number of craters larger than 1 km in diameter produced on the moon per $\mathrm{km^2}$ has been estimated as

\begin{equation}
N_l(T) = a_l \ \left[\exp{(\lambda_l \ T)}-1 \right] + b_l \ T
\label{eq_lcum}
\end{equation}

\noindent where $T$ is the time measured backwards from the present in Ga \citep{Neukum2001SSRv}.  From \citet{Neukum2001SSRv}, the coefficients are $a_l = 5.44 \times 10^{-14} \ \mathrm{km^{-2}}$, $b_l = 8.38 \times 10^{-4} \ \mathrm{km^{-2} \ \mathrm{Ga}^{-1}}$, $\lambda_l = 6.93 \ \mathrm{Ga^{-1}}$ ($\tau_l = 1/\lambda_l = 0.144$ Ga).  \citet{Marchi2009AJ} find slightly different, but still similar values of $a_l = 1.23 \times 10^{-15}  \ \mathrm{km^{-2}}$, $b_l = 1.30 \times 10^{-3} \ \mathrm{km^{-2} \ \mathrm{Ga}^{-1}}$, $\lambda_l = 7.85 \ \mathrm{Ga^{-1}}$ ($\tau_l = 1/\lambda_l = 0.127$ Ga).  This chronology is based on crater counts of areas for which radiometrically-dated samples are available (although as discussed in Sec.~\ref{sec_bg} there is some debate over the actual source regions of some of the samples), and is not constrained prior to $\sim$4.1 Ga even though it is often plotted extending back to 4.5 Ga.  Recent model-based chronologies for the moon \citep[eg.][]{Morbidelli2012EPSL} are in general agreement with this chronology from $\sim$4.1 Ga to the present time.

Taking the derivative of Eq.~\ref{eq_lcum} gives the differential production rate of craters larger than 1 km diameter per $\mathrm{km^2}$ per Ga

\begin{equation}
\frac{\mathrm{d}N_l}{\mathrm{d}T} = a_l \ \lambda_l \ \exp{(\lambda_l \ T)} + b_l
\label{eq_ldiff}
\end{equation}

\noindent Eq.~\ref{eq_ldiff} can be normalized to give a value of 1 at present time ($T=0$), which for the generally-satisfied case where $a_l \ \lambda_l \ll b_l$ is


\begin{equation}
\frac{\mathrm{d}N_l^*}{\mathrm{d}T} = \frac{a_l \ \lambda_l}{b_l} \  \exp{(\lambda_l \ T)} + 1
\label{eq_ldiffn}
\end{equation}

\noindent The asterisk denotes the normalized form.  The normalized form of the cumulative expression (Eq.~\ref{eq_lcum}) can be found either by integrating Eq.~\ref{eq_ldiffn} or by dividing Eq.~\ref{eq_lcum} through by $b_l$

\begin{equation}
N_l^*(T) = \frac{a_l}{b_l} \ \left[\exp{(\lambda_l \ T)}-1 \right] + T
\label{eq_lcumn}
\end{equation}

\noindent Note that the resulting expression has units of time.  The intuitive interpretation is that if the cratering rate varies in time and a given number of craters has accumulated in time $T$, $N^*(T)$ is the amount of time it would have taken to accumulate the same number of craters if the impact rate were fixed at the present value.  For an impact rate that increases at earlier times, it will always be the case that $N^*(T) \geq T$.  Normalized forms such as this will be useful in the subsequent sections, as they can be multiplied by the crater production function at the current time to give the total cumulative crater production for any value of T.

\citet{Schmedemann2014PSS} assume that the lunar curve can be directly scaled to Vesta.  If $f$ is the current rate of formation of craters 1 km and larger on Vesta (per $\mathrm{km^2}$ per Ga) and $r$ is the ratio of the formation rate on Vesta to the rate on the Moon, then $r = f/b_l$ and Eq.~\ref{eq_lcum} is modified to give a `lunar-like' curve

\begin{equation}
N_{ll}(T) = r \ N_l(T) = a_l \ r \ \left[\exp{(\lambda_l \ T)}-1 \right] + f \ T
\label{eq_llcum}
\end{equation}

\noindent \citet{Schmedemann2014PSS} find that for Vesta, $f = 0.01979 \ \mathrm{km^{-2} \ Ga^{-1}}$ and $r = 23.62$.  Taking the derivative of Eq.~\ref{eq_llcum} (or alternatively multiplying Eq.~\ref{eq_ldiff} by $r$) gives the differential production rate of craters 1 km and larger per $\mathrm{km^2}$ per Ga on Vesta

\begin{equation}
\frac{\mathrm{d}N_{ll}}{\mathrm{d}T} = r \ \frac{\mathrm{d}N_{l}}{\mathrm{d}T} = a_l \ \lambda_l \ r \ \exp{(\lambda_l \ T)} + f
\label{eq_lldiff}
\end{equation}

\noindent Eq.~\ref{eq_lldiff} can be normalized to give a value of 1 at present time ($T=0$).  This yields the same result as the normalized lunar curve (Eq.~\ref{eq_ldiffn}), as expected since they are just scaled versions of each another

\begin{equation}
\frac{\mathrm{d}N_{ll}^*}{\mathrm{d}T} = \frac{a_l \ \lambda_l}{b_l} \  \exp{(\lambda_l \ T)} + 1
\label{eq_lldiffn}
\end{equation}

\noindent Similarly, the normalized cumulative curve is

\begin{equation}
N_{ll}^*(T) = \frac{a_l}{b_l} \ \left[\exp{(\lambda_l \ T)}-1 \right] + T
\label{eq_llcumn}
\end{equation}

As discussed in Sec.~\ref{sec_bg_sub1}, while it may be reasonable to scale the impact rate and chronology function between bodies in the inner Solar System, scaling it to the asteroid belt does not have the same physical basis, and in fact it may imply a particular history for the asteroid belt that may not be plausible.  The simplest physical interpretation of taking the chronology function inferred for the moon and simply scaling it to match the current impact rate in the asteroid belt is that the asteroids are the primary impactors on other asteroids (which is most likely true), but also that the impact rate within the belt going back in time, and hence its total mass, directly tracks the curve given by Eq.~\ref{eq_lldiff}.

\section{A Model Chronology Curve Based on Main-Belt Dynamics}
\label{sec_m}

Here we present a chronology curve for the asteroid belt based on the recent dynamical results described in Sec.~\ref{sec_bg}, assuming three main processes: (1) A primordial depletion from an initial impacting mass $M_o$ (in units of current asteroid belt mass) with timescale $\tau_{pd}$, which could be due to the depletion in mass of the primordial asteroid belt itself, perhaps by embedded planetary embryos \citep{Wetherill1992Icar, Petit2001Icar, OBrien2007Icar}, or alternatively by the decay of scattered leftover planetesimals from the terrestrial planet region; (2) Rapid loss of mass by a factor $f_{LHB} \sim 2$ starting at time $T_{LHB}$ with decay time constant $\lambda_{LHB}$, triggered by the sweeping of resonances through the belt during a rapid phase of planetesimal-driven giant planet migration \citep[ie.~the Nice Model,][]{Gomes2005Nat, Morbidelli2010AJ}; and (3) Loss of a factor $f_{cd} \sim 2$ by post-LHB chaotic diffusion \citep{Minton2010Icar}.  We use the subscript LHB here for the time of the instability, as it coincides with the Late Heavy Bombardment on the terrestrial planets.

One key parameter is the time of destabilization of the asteroid belt by the Nice Model resonance sweeping event $T_{LHB}$, which is often quoted as being around 3.9 Ga based on early dating of lunar samples \citep[eg.][]{Tera1974EPSL}, but more recent work \citep[eg.][]{Bottke2012Nat, Morbidelli2012EPSL, Marchi2013NatGSci} suggests is more likely $\sim$4.1 Ga.  We nominally assume that $\lambda_{LHB} = \lambda_l$ (ie.~that the decay follows the same profile as the lunar curve), although this does not necessarily have to be the case.  Two other important parameters are the original mass of impactors in or otherwise affecting the asteroid belt $M_o$ (relative to the current mass), and the timescale of the decay of the primordial impact rate $\tau_{pd}$.  We nominally use a value of $\tau_{pd}$ = 25 Ma based on the results of \citet{Bottke2005bIcar}, who find that the primordial impact flux in the asteroid belt dropped to $\sim$2\% of the initial value within $\sim$100 Ma of its formation, roughly consistent with an exponential decay timescale of 25 Ma.  $M_o$ is variously estimated to be on the order of hundreds to thousands \citep[see, eg.,][]{Weidenschilling1977ApSS,Wetherill1992Icar, Petit2001Icar, Bottke2005bIcar, OBrien2007Icar}.  In Sec.~\ref{sec_apply} we constrain that value, using the record of large impact basins on the surface of Vesta.

\subsection{Derivation}
\label{sec_m_deriv}

\citet{Minton2010Icar} find that there has likely been a decay in the number of asteroids in the main belt due to chaotic diffusion following the LHB resonance sweeping event, which is when the current dynamical structure of the asteroid belt is assumed to have been established.  The parameterization of the decay assumed here is based on \citeauthor{Minton2010Icar}, Table 1 and Eq.~4, although we rescale it to give relative number in the belt ($n_{cd}(T)$, equal to 1 at the present time), rather than fraction remaining, and we parameterize it in Ga rather than years.


\begin{equation}
n_{cd}(T) = C_{cd} \ ((T_{LHB}-T)/(1 \ \mathrm{Ga})+0.001)^{-0.0834}
\label{eq_ncd}
\end{equation}

\noindent where

\begin{equation}
C_{cd} = (T_{LHB}/(1 \ \mathrm{Ga})+0.001)^{0.0834}
\label{eq_Ccd}
\end{equation}

\noindent The loss factor due to chaotic diffusion since $T = T_{LHB}$ is then

\begin{equation}
f_{cd} = C_{cd} \ (0.001)^{-0.0834} \sim 2
\end{equation}

\noindent Note that this is probably a lower limit to the loss that may occur, as we do not explicitly include the loss that may be due to collisional grinding. \citet{Durda1998Icar} find that there could be as much as a factor of 3 loss due to collisional grinding.  \citet{Bottke2005bIcar} found that there would be `modest' depletion due to collisional grinding.  While they did not quantify the amount, it would likely be less than found by \citeauthor{Durda1998Icar},  since \citeauthor{Durda1998Icar} assume a particularly weak strength law for asteroids.

To get the normalized differential cratering curves, we make two simplifying assumptions:~1) The impact rate in the asteroid belt is directly proportional to the total mass of impactors; and 2) The impact velocity is constant with time.  These together imply that the production rate of a crater of a given size is directly proportional to the mass of impactors of a given size.  The first assumption is not strictly true given that the mass of the asteroid belt is dominated by the largest bodies and while the loss of one of them may significantly change the belt mass, it would have little effect on the population of small impactors.  Thus, when we talk about the mass of impactors, we refer to an idealized distribution that smooths over stochastic variations at the large size end.  \citet{Marchi2013NatGSci} find that while the E-belt impactors would have a higher impact velocity than main-belt asteroids, their numbers are relatively small compared to main-belt impactors, in part justifying our second assumption.  It is possible that scattered planetesimals from the terrestrial planet zone, however, would have a different impact velocity and collision probability with the asteroids than other main-belt impactors.  As the relative contributions of the different impacting populations at different times are not well-constrained, we keep with assumptions 1 and 2 in the derivations that follow.

We can get a normalized differential cratering curve for $T$ less than $T_{LHB}$ by combining the chaotic diffusion term (Eq.~\ref{eq_ncd}) with an exponential decay term for the depletion of the belt following the instability at $T_{LHB}$

\begin{equation}
\frac{\mathrm{d}N_{m<}^*}{\mathrm{d}T} = C_{LHB} \ \exp{(\lambda_{LHB} \ T)} + C_{cd} \ ((T_{LHB}-T)/(1 \ \mathrm{Ga})+0.001)^{-0.0834}
\label{eq_mdiffnl}
\end{equation}

\noindent where $\lambda_{LHB}$ may be similar to the exponential decay constant $\lambda_l$ for the \citet{Neukum2001SSRv} curve, although not necessarily.  The coefficient $C_{LHB}$ of the LHB decay term is obtained by setting $C_{LHB} \ \exp{(\lambda_{LHB} \ T)} = (f_{LHB} \ f_{cd} - f_{cd})$ at $T = T_{LHB}$, which sets the belt mass at $T = T_{LHB}$ to be $f_{LHB} \ f_{cd}$ times the present mass

\begin{equation}
C_{LHB} = (f_{LHB} \ f_{cd} - f_{cd}) / \exp{(\lambda_{LHB} \ T_{LHB})}
\label{eq_clhb}
\end{equation}

\noindent For $T$ greater than $T_{LHB}$, the normalized differential curve is based on the normalized belt mass at $T = T_{LHB}$ (which is $f_{LHB} \ f_{cd}$) combined with a primordial depletion term to account for the decrease in bombardment rate immediately following the formation of the Solar System

\begin{equation}
\frac{\mathrm{d}N_{m>}^*}{\mathrm{d}T} = f_{LHB} \ f_{cd} + C_{pd} \ \exp{((T-T_o)/\tau_{pd})}
\label{eq_mdiffng}
\end{equation}

\noindent where 

\begin{equation}
C_{pd} = (M_o - f_{LHB} \ f_{cd})
\label{eq_cpd}
\end{equation}

\noindent and $T_o$ is the time at which the bombardment of Vesta's surface begins to be recorded.  The first solids in the Solar System formed at 4.567 to 4.568 Ga \citep{Bouvier2010NatGSci, Connelly2012Sci} and the excitation of potential impacting bodies likely did not begin until the gas disk dissipated several Ma after that \citep{Haisch2001ApJ, Kita2005ASPC}.  Similarly, it may have taken at least several Ma for Vesta to differentiate and form its crust \citep[see][for a review]{McSween2011SSRv}.  Hence, we nominally set $T_o$ = 4.56 Ga.

While the exponential form of the primordial decay term in Eq.~\ref{eq_mdiffng} is straightforward and involves the fewest free parameters, an alternative parameterization using a `stretched' exponential function \citep[eg.][]{Dobrovolskis2007Icar} that that includes an additional free parameter to give a longer `tail' to the decay is described in \ref{sec_appendix}.  This may prove useful in future work that attempts a more detailed fitting of the model presented here to the results of numerical simulations of possible impactor populations.

To obtain the cumulative chronology curves, we integrate Eqns.~\ref{eq_mdiffnl} and \ref{eq_mdiffng}.  For $T$ less than $T_{LHB}$, the integral of Eq.~\ref{eq_mdiffnl} gives
 
\begin{align}
N_{m<}^*(T) = & \int_0^T \frac{\mathrm{d}N_{m<}^*}{\mathrm{d}T'} \ \mathrm{d}T' \nonumber \\
= & \ \frac{C_{LHB}}{\lambda_{LHB}} \ \left[\exp{(\lambda_{LHB} \ T)}-1 \right] \ - \nonumber \\
& \ \frac{C_{cd}}{0.9166 \ \mathrm{Ga^{-1}}} \ ((T_{LHB}-T)/(1 \ \mathrm{Ga})+0.001)^{0.9166} + C_1
\label{eq_mcumln}
\end{align}

\noindent where

\begin{equation}
C_1 = \frac{C_{cd}}{0.9166 \ \mathrm{Ga^{-1}}} \ (T_{LHB}/(1 \ \mathrm{Ga})+0.001)^{0.9166}
\label{eq_c1}
\end{equation}

\noindent For $T$ greater than $T_{LHB}$, the integral of Eqns.~\ref{eq_mdiffnl} and \ref{eq_mdiffng} gives

\begin{align}
N_{m>}^*(T) = & \int_0^{T_{LHB}} \frac{\mathrm{d}N_{m<}^*}{\mathrm{d}T'} \ \mathrm{d}T' + \int_{T_{LHB}}^T \frac{\mathrm{d}N_{m>}^*}{\mathrm{d}T'} \  \mathrm{d}T' \nonumber \\
= & \ C_2 + f_{LHB} \ f_{cd} \ (T-T_{LHB}) \ + \nonumber \\
& \ C_{pd} \ \tau_{pd} \ \exp{((T-T_o)/\tau_{pd})} - C_3
\label{eq_mcumgn}
\end{align}

\noindent where

\begin{equation}
C_2 = \frac{C_{LHB}}{\lambda_{LHB}} \ \left[\exp{(\lambda_{LHB} \ T_{LHB})}-1 \right] - \frac{C_{cd}}{0.9166 \ \mathrm{Ga^{-1}}} \ (0.001)^{0.9166} + C_1
\label{eq_c2}
\end{equation}

\noindent and 

\begin{equation}
C_3 = C_{pd} \ \tau_{pd} \ \exp{((T_{LHB}-T_o)/\tau_{pd})}
\label{eq_c3}
\end{equation}

\noindent Note that $C_2$ is equal to Eq.~\ref{eq_mcumln} evaluated at $T=T_{LHB}$.

\section{Crater Production Functions}
\label{sec_prodfunc}

The crater production function, which we denote here as $F(D)$, gives the number of craters of a given size $D$ or larger per unit area per unit time (here we use units of $\mathrm{km^{-2} \ \mathrm{Ga}^{-1}}$).  $N^*(T)$ from Eqns.~\ref{eq_mcumln} and \ref{eq_mcumgn} (for the model chronology) or Eq.~\ref{eq_llcumn} (for the lunar-like chronology) can be used to scale the crater production function to give the cumulative crater production for any time $T$.  The cumulative number of craters of a given diameter or larger produced per unit area since time $T$, where $T=0$ is the present time, is found by

\begin{equation}
N(D,T) = N^*(T) \ F(D)
\label{eq_ncum_abs}
\end{equation}

\noindent For $F(D)$ on Vesta we use a model production function $F_{m}(D)$ that is derived using the model main belt size distribution from \citet{Bottke2005bIcar}, which is constrained by the observed main-belt size distribution at large sizes and a range of other constraints such as the cosmic-ray exposure ages of meteorites and the number of asteroid families.  Crater scaling laws \citep{Holsapple2007Icar} and estimates of the main-belt impact rate are used to convert this main-belt size distribution into a crater production function for Vesta, as outlined in \citet{Marchi2010PSS, Marchi2012PSS, Marchi2012Sci, Marchi2014PSS}.  \citet{Schmedemann2014PSS} take a somewhat different approach by scaling the estimated lunar crater production function \citep[from][]{Neukum2001SSRv} to Vesta, accounting for the differences in impact velocity and the relative numbers of possible impactors.  While there are some differences between these two production functions, they both give similar  production rates of 1 km-scale craters.  We will show in Sec.~\ref{sec_apply} that the primary reason for different age estimates of older terrains on Vesta lies in the differences between the chronology functions, not the assumed crater production function.

\section{Applying the Model Chronology to Vesta}
\label{sec_apply}

From the crater catalog of \citet{Marchi2012Sci}, with recent updates to include the north polar region, there are five craters roughly 200 km diameter and larger and nine craters roughly 100 km diameter and larger.  Other unpublished crater catalogs have been compiled by members of the Dawn team, with the number of craters 100 km in diameter and larger ranging from 6 to 11, and it is possible that numerous craters of that size were erased over Vesta's history, especially by the formation of the large Rheasilvia and Veneneia basins.  Since the crater SFD of all but the two largest craters may have been influenced to some degree by erasure process and therefore may not accurately reflect the production function, we only use these two largest craters to constrain the parameters in the expressions for the chronology curve, namely $M_o$.

Figure \ref{fig_prod_func_fit} shows the result of fitting the model production function $F_{m}(D)$ to the two largest craters on Vesta (note that while craters down to $\sim$100 km diameter are shown, they are not included in the fit).  We solve for the $N^*$ value such that $N^* \ F_{m}(D)$ best matches the two large craters, giving an $N^*$ value of 27.4 Ga.  In addition, we perform the same calculation assuming that the expected value of craters of that size over Vesta's lifetime is either 1 or 4 (approximately a 1-sigma range).  This gives $N^*$ values of 13.7 and 54.8 Ga, respectively.  Figure \ref{fig_prod_func_fit} shows that, based on the model production function assumed here, it is possible that numerous $\sim$200 km diameter craters could have formed and been erased over Vesta's history, and it is highly likely that many $\sim$100 km diameter craters have been erased.

As described in Sec.~\ref{sec_m}, many of the parameters in Eqns.~\ref{eq_mcumln} and \ref{eq_mcumgn} can be estimated from modeling or theory, and we summarize those values in Table \ref{table_param}.  The effects of varying those parameters from their nominal values will be described later in this section.  To estimate $M_o$, the original mass of impactors relative to the current mass of the main belt, we can solve Eqn.~\ref{eq_mcumgn}, the cumulative model chronology function, for $M_o$ by setting Eq.~\ref{eq_mcumgn} equal to the $N^*$ values determined by the crater count fits in Fig.~\ref{fig_prod_func_fit} at $T = T_o$.  This gives $M_o$ = 836 for the model production function, with a 1-sigma range of 288 to 1932.  These $M_o$ values assume a timescale for primordial depletion $\tau_{pd}$ of 25 Ma, and could be larger if $\tau_{pd}$ is smaller.  The current mass of the main belt is estimated to be roughly 0.0006 Earth masses \citep{Krasinsky2002Icar}, so these $M_o$ values correspond to a possible range of 0.17 to 1.16 Earth masses for the primordial impacting mass, consistent with estimates of amount of mass in the primordial asteroid belt \citep[eg.,][]{Weidenschilling1977ApSS, Wetherill1992Icar, Petit2001Icar, Bottke2005bIcar, OBrien2007Icar}.

Figure \ref{fig_diff} shows a comparison of the normalized differential production rate of craters $\mathrm{d}N^*/\mathrm{d}T$ in the lunar-like chronology (Eq.~\ref{eq_lldiffn} from Sec.~\ref{sec_lun}) and the model chronology (Eqns.~\ref{eq_mdiffnl} and \ref{eq_mdiffng} from Sec.~\ref{sec_m}), for the nominal parameter values given above.  Also shown is a constant linear production rate curve.  Note that the lunar-like curve prior to 4.1 Ga is an extrapolation, since the lunar chronology on which it is based is only constrained back to $\sim$4.1 Ga.  While the lunar-like curve increases rapidly prior to $\sim$3 Ga, the model curve first increases to $\sim$4 times its current value prior to 4.1 Ga (the time of destabilization of the asteroid belt and E-Belt), then increases again going back to $T_o$, due to the larger primordial mass of the asteroid belt, and/or scattered planetesimals from the terrestrial planet region that can strike the asteroids.  For any given time prior to $\sim$ 3 Ga, the lunar-like curve implies a much larger impact rate in the asteroid belt than the model curve.

The integrals of the differential curves from Fig.~\ref{fig_diff} give the normalized chronology functions $N^*(T)$. Figure \ref{fig_cum} shows the a comparison of the normalized chronology functions $N^*(T)$ in the lunar-like chronology (Eq.~\ref{eq_llcumn} from Sec.~\ref{sec_lun}) and the model chronology (Eqns.~\ref{eq_mcumln} and \ref{eq_mcumgn} from Sec.~\ref{sec_m}), for the nominal parameter values given above, along with a linear chronology curve.  Note that while the model differential curve in Fig.~\ref{fig_diff} shows sharp increases near T = 4 and 4.5 Ga, the corresponding increases in the cumulative curve are more gradual, because the differential curves are integrated from T = 0 back to the time in question rather than just over a small interval where the changes occur.  The second plot in Fig.~\ref{fig_cum} shows an approximately 1-sigma range based on uncertainties in the estimate of the initial impacting mass $M_o$.  

Figure \ref{fig_n1km} shows the chronology curves for Vesta in terms of absolute numbers of craters larger than 1 km, using Eq.~\ref{eq_ncum_abs} along with the normalized $N^*(T)$ curves from Fig.~\ref{fig_cum} and the crater production function $F_{m}(D)$.  The lunar-like curve of \citet{Schmedemann2014PSS} (our Eq.~\ref{eq_llcum}) is also shown.  The two curves are roughly the same for $T<3.5$ Ga, since the production rate of 1 km craters is roughly the same in the lunar-like and model production functions.  However, they diverge significantly prior to 3.5 Ga, because of the divergence of the chronology functions (Fig.~\ref{fig_cum}).  For any given time prior to $\sim$ 3.5 Ga, the lunar-like curve implies that a given surface of that age would have a higher crater density than is implied by the model chronology (or alternatively, the lunar-like chronology gives a younger age than the model chronology for a surface with a given crater density).

For the lunar-like curve in Fig.~\ref{fig_n1km}, crater densities as high as predicted around $\sim$4.1 Ga would not actually be possible, as the surface would become saturated and the actual observed crater numbers would lie below the production curve \citep[eg.][]{Gault1970RadSci}.  For the model chronology curves, it is possible that the levels achieved near $T = T_o$ would be close to the empirical saturation level as well, and that could potentially affect the conversion of crater density to absolute age.  This is discussed in further detail in \citet{Marchi2012Sci}.

It is illustrative to compare our model chronology to the lunar chronology.  Figure \ref{fig_vestluncomp} shows the crater production rates and cumulative chronology curves in terms of 1 km diameter craters for both the moon and Vesta, where the lunar curves are given by Eqns.~\ref{eq_lcum} and \ref{eq_ldiff} and the Vesta curves are given by Eqns.~\ref{eq_mdiffnl}, \ref{eq_mdiffng}, \ref{eq_mcumln} and \ref{eq_mcumgn}, scaled appropriately by the production rate of 1 km diameter craters from the Vesta model production function $F_{m}(D)$ (approx.~0.0190 $\mathrm{km^{-2} \ \mathrm{Ga}^{-1}}$).  Note in particular that while the impact rate curve for Vesta lies above the lunar curve by a factor of $\sim$20 for $T<3$ Ga, it is about a factor of $\sim$10 lower around 4 Ga, the nominal time of the LHB.  The reason these curves can follow substantially different profiles (rather than simply being scaled versions of one another, was briefly alluded to in Sec.~\ref{sec_bg_sub1}.  Around 4.1 Ga in our model chronology, the asteroid belt is depleted by a factor of $\sim$4 through a combination of resonance sweeping and subsequent chaotic diffusion of bodies out of the belt.  While this only leads to a decrease in impact rate of factor of $\sim$4 within the asteroid belt (and hence on Vesta), the several asteroid belts worth of mass that are rapidly ejected from the asteroid belt lead to a huge increase in the impact rate on the moon and terrestrial planets.  For $T<3$ Ga, after this influx of material has decayed, impacts on the moon occur at a much lower rate, set by the rate at which bodies slowly leak out of the asteroid belt to become NEAs.  While we do not explicitly calculate the lunar impact rate from the dynamical assumptions of our model chronology, we note that \citet{Bottke2012Nat} and \citet{Morbidelli2012EPSL} have shown that the impact rate on the moon implied by the dynamical scenario on which we build our model chronology is generally consistent with that of \citet{Neukum2001SSRv} over the last 4.1 Ga, which is what we plot for the lunar curves in Fig.~\ref{fig_vestluncomp}.

Figure~\ref{fig_cum_range} shows how the model chronology curve is affected by varying several of the key parameters in Eqns.~\ref{eq_mcumln} and \ref{eq_mcumgn} from their nominal values given in Table \ref{table_param}.  $f_{LHB}$ is varied from 1 to 10, relative to its nominal value of 2.  With $f_{LHB}$ = 1, there is no depletion and hence no change in the impact rate at $t_{LHB}$, the impact rate only changes during the early primordial depletion phase.  $f_{LHB}$ = 10 is consistent with the early Nice Model simulations \citep{Gomes2005Nat}, although subsequent models have revised that value downward to the current estimate of $\sim$2 \citep{Morbidelli2010AJ}.  The primordial depletion timescale $\tau_{pd}$ is varied by a factor of 2 around its nominal value of 25 Ma (ie. a range of 12.5 Ma to 50 Ma).  In both cases, the range of variation in the initial $N^*(T)$ value at $T_o$ = 4.56 Ga is comparable to the range of variation due to the 1-sigma range in $M_o$ estimates, although the shape of the curves may be significantly affected (particularly in the case of high $f_{LHB}$).

Figure \ref{fig_vesta_cc} shows our model chronology function applied to actual crater counts on Vesta.  Two regions are chosen, the floor of Rheasilvia, and the highly-cratered terrains (HCTs) in the northern hemisphere identified by \citet{Marchi2012Sci} and subsequently revised.  The model production function $F_{m}(D)$ is fit to the crater counts for these two regions, such that $N^* \ F_{m}(D)$ best matches the observed crater size-frequency distributions.  The resulting $N^*$ value can be related to the surface age with the model chronology function in Fig.~\ref{fig_cum} (where the model chronology function is from Eqns.~\ref{eq_mcumln} and \ref{eq_mcumgn}, with the values in Table \ref{table_param} and $M_o$ = 836).

We find an age of Rheasilvia of $\sim$1 Ga, consistent with the earlier estimates by \citet{Marchi2012Sci}.  Assuming that the craters on the HCT regions are primary and fitting to the small end ($D<8$ km) of the crater SFD, we find an age of $\sim$4.3 Ga, consistent with them being amongst the oldest terrains on Vesta.  It is likely that the HCTs are saturated or close to saturation, so the true age could be closer to $\sim$4.5 Ga.  Because the larger craters on the HCTs lie below the production curve while smaller craters are saturated, it has been suggested that the impacting population on the HCT regions was primarily a steep secondary population from the Rheasilvia impact.  However, more detailed analysis of those regions finds that few craters in that area show morphological signs of being secondaries (which would have formed at less than Vesta's escape velocity of $\sim$350 m/s).  The most likely explanation is that the HCTs are not really a specific geological unit, but are regions chosen specifically because they have the highest crater density, and the model production function in the $\sim$3-20 km range of craters on the HCTs has a slope somewhat shallower than -2.  With a slope shallower than -2, erasure by large craters dominates and the crater population at smaller sizes mirrors the slope of the production population, with the overall crater density dependent on stochastic erasure by large craters \citep[eg.][]{Chapman1986Sats, Melosh1989ImpCrat, Richardson2009Icar}.  In that case, the regions chosen as the HCTs, with the highest density of small craters, will be the ones that avoided experiencing many larger impacts, hence the low crater counts for $D>8$ km.

\section{Discussion}
\label{sec_discuss}

We present a cratering chronology for Vesta based on the best current understanding of the dynamical evolution of asteroid belt and the impact history of the inner Solar System, and we calibrate it to Vesta using the record of large craters on its surface.  In this chronology, Vesta would have experienced two main stages:~1) An early declining impact flux due to the dynamical depletion of the asteroid belt and/or scattered planetesimals from the terrestrial planet region; and 2) A secondary decrease in impact flux around $\sim$4 Ga due to the sweeping of resonances through the asteroid belt driven by giant planet migration, as well as the chaotic diffusion of unstable asteroids out of the main belt.  While there are necessarily some simplifications and uncertainties in the model, our chronology is currently the best `educated guess' we have as to Vesta's impact history, and is broadly consistent with estimates of the initial amount of mass present in the asteroid belt and with Vesta having an ancient cratered surface.

The meteorite record provides additional evidence for the collisional and dynamical history of the asteroid belt, and of Vesta.  The HED meteorites from Vesta, as well as the H chondrites, show a broad peak in impact-reset ages from $\sim$3 to 4.2 Ga and a relative lack of ages between $\sim$4.2 and 4.5 Ga \citep{Bogard1995MAPS, Bogard2003MAPS, Swindle2009MAPS, Bogard2011CDE}.  \citet{Marchi2013NatGSci} show that this is likely a consequence of the destabilization of the `E-Belt,' located at the inner edge of the primordial asteroid belt.  While those bodies would not lead to a significantly increased impact rate in the asteroid belt, they would lead to an increase in high-velocity impacts, which would be capable of resetting Ar-Ar ages on asteroid surfaces.  Those surfaces would be sampled by later impacts that would eject material that could eventually evolve onto Earth-crossing orbits.  In the case of Vesta, Rheasilvia is the likely candidate, as it appears to be the last major impact, and could have sampled material from numerous pre-existing impact basins with a range of ages.  Our chronology, which includes the destabilization of the E-Belt and asteroid belt at $\sim$4.1 Ga, with a plateau in the impact rate prior to that leading back to the primordial spike in the impact rate, is consistent with this picture.

We find an age of $\sim$1 Ga for Rheasilvia using our model chronology, consistent with previous estimates \citep{Marchi2012Sci}.  This is consistent with several other lines of evidence as well.  Rheasilvia has a quite fresh appearance relative to all other ancient basins.  The size distribution of the Vesta family members, or `vestoids,' is quite steep compared to the background population. \citet{Marzari1996AA, Marzari1999Icar} showed that its size distribution would collisionally grind down if the family were older than $\sim$1 Ga, suggesting its relatively recent formation.  A young age is also consistent with the much lower abundance of Hydrogen within Rheasilvia Basin compared to the rest of the surface, as found by the Gamma Ray and Neutron Detector (GRaND) on Dawn \citep{Prettyman2012Sci} and interpreted as exogenic in origin.  \citet{Nesvorny2008Icar} estimate that the Vesta family would have to be $\sim$1-2 Ga old in order to explain the presence of `fugitive' vestoids that lie outside the bounds of the main family, although if radiation forces like the YORP effect that can lead to a non-random distribution of spin axes are taken into account, the age is closer to 1 Ga (D.~Nesvorny, pers.~comm.).  It may seem strange that the largest crater on Vesta would be so young, where we define young here as forming in the last $\sim$3 Ga, during which time the impact rate is roughly the same as the current rate.  We find that while it may be a low-probability event, it is still plausible given the impact history implied by the model chronology.  Assuming that one Rheasilvia-sized basin formed over the lifetime of Vesta, the probability that it formed in the last 3 Ga is roughly estimated as $N^*(T=3 \ \mathrm{Ga})/N^*(T=T_o)$ = 0.134, or 13.4\%, assuming the nominal model chronology parameters.

As described in Sec.~\ref{sec_bg_sub1}, we do not account for several possible additional sources of impactors, including cometary influx during the LHB \citep{Gomes2005Nat, Levison2009Nat}, or scattered planetesimals during the formation of Jupiter \citep{Turrini2014PSS}, and in \ref{sec_m_deriv} we assume that the impact rate on Vesta is directly proportional to the total impacting mass and that the impact velocity is constant.  We also chose a value of $\tau_{pd}$ for the primordial decay of the impact flux of 25 Ma, based on models of the depletion of the asteroid belt through perturbations by primordial planetary embryos.  If the depletion occurred through a different process (eg.~the `Grand Tack' of \citet{Walsh2011Nat}), the depletion would have been faster, and the resulting primordial impacting mass $M_o$ in our model would have to be correspondingly higher.  While all of these are reasonable assumptions to make, given the uncertainties involved, it does mean that the chronology presented here should not be taken as the final word for calculating ages on Vesta's surface, and should certainly not be used to calculate 3-significant-figure ages as is common in the crater counting literature.  Rather, we provide a general framework for Vestan chronology that can be refined as further constraints may become available, and which can be used to provide reasonable age estimates in the context of our current understanding of Solar System evolution.  Because of the uncertainties involved, we stress that in presenting crater counts for Vesta, the age should not be the only piece of information given.  Rather, since any absolute age calculation is model-dependent, a quantitative measure of the crater density (such as the number of craters larger than a certain size) should be given, as well as any other relevant information such as the assumed impacting population, so that the reader can do their own comparison between the results of different groups who may be making different assumptions.

We contrast our approach with that of \citet{Schmedemann2014PSS}, who apply the lunar chronology curve of \citet{Neukum2001SSRv} to Vesta by scaling it to Vesta's current impact rate in the main belt.  As discussed in Secs.~\ref{sec_bg_sub1} and \ref{sec_lun}, while it may be reasonable to scale the curve between different bodies in the inner Solar System since they are all impacted by a common population, the NEAs, it is not necessarily true that one can simply scale it to the asteroid belt.  The dynamical history that delivers the impactors from the main belt to the NEA population may imply significantly different collisional histories for bodies in the asteroid belt compared to the moon and terrestrial planets.  Fig.~\ref{fig_vestluncomp} illustrates this clearly, as the Lunar curve shown there is broadly consistent (at least back to 4.1 Ga) with the dynamical scenario \citep[eg.][]{Bottke2012Nat, Morbidelli2012EPSL} that is the basis of our model chronology curve for Vesta, yet the two chronology curves differ significantly from one another.

Taking the chronology function inferred for the moon and scaling it to match the current impact rate in the asteroid belt, as done by \citet{Schmedemann2014PSS}, is not based on any specific physical scenario, but the simplest physical interpretation of such a scaling is that the primary impactors in the asteroid belt are other asteroids (which is likely true), and that the impact rate in the belt going back in time, and hence its total mass, directly tracks the curve given by Eq.~\ref{eq_lldiff} (shown in Fig.~\ref{fig_diff}).  While lunar chronology itself is not constrained prior to 4.1 Ga, the scaling used to give the lunar-like chronology for Vesta may imply an impact rate that continues to follow the dashed curve in Fig.~\ref{fig_diff} prior to 4.1 Ga, possibly giving an unreasonably large primordial impact rate and the production of significantly more large basins that the two seen on Vesta today.  While there are potentially other scenarios that could avoid this issue, we caution that the full implications and physical interpretation of using a lunar-like chronology in the asteroid belt should be more thoroughly explored if it is going to be employed in future work.

Dawn will arrive at Ceres in 2015, giving opportunity to test our model chronology and perhaps refine it with further constraints, such as the record of large impact basins on Ceres.  However, Ceres is likely a very different body than Vesta and may pose several difficulties for the interpretation of its cratering record.  Given that Ceres may contain a significant fraction of water ice \citep[eg.][]{McCord2011SSRv}, the scaling law for crater formation will likely differ from that of Vesta, although that can be at least partially accounted for.  More problematic is that, as discussed by \citet{Bland2013Icar}, significant water ice could mean that many craters, especially those in the warmer equatorial regions, could viscously relax on short timescales, leaving little or no record of their existence.

\section*{Acknowledgments}

D.~P.~O'Brien is supported by grant NNX10AR21G from NASA's Dawn at Vesta Participating Scientist Program.  The Dawn mission to asteroid Vesta and dwarf planet Ceres is managed by JPL, for NASA's Science Mission Directorate, Washington, DC.  UCLA is responsible for overall Dawn mission science.  We thank David Minton and an anonymous referee for their helpful reviews and comments.

\appendix

\section{Alternative Parameterization of Primordial Depletion}
\label{sec_appendix}

An alternative parameterization of the primordial decay uses a `stretched' exponential function \citep[eg.][]{Dobrovolskis2007Icar} with a decay timescale $\tau_{pd}$ modified by an exponent $\beta_{pd}$ less that 1 that gives a longer `tail' to the decay.  While we do not use this form here, we include it for completeness, as it may provide useful for fitting the model presented here to the results of numerical simulations of the primordial decay of the impactor population.

Using this new form, Eq.~\ref{eq_mdiffng} for the normalized differential curve for $T$ greater than $T_{LHB}$ becomes

\begin{equation}
\frac{\mathrm{d}N_{m>}^*}{\mathrm{d}T} = f_{LHB} \ f_{cd} + C_{pd} \ \exp{\left(-\left(\frac{T_o-T}{\tau_{pd}}\right)^{\beta_{pd}}\right)}
\label{eq_mdiffng_new}
\end{equation}

\noindent For $T$ greater than $T_{LHB}$, the integral of Eqns.~\ref{eq_mdiffnl} and \ref{eq_mdiffng_new} gives the new normalized cumulative curve

\begin{align}
N_{m>}^*(T) = & \int_0^{T_{LHB}} \frac{\mathrm{d}N_{m<}^*}{\mathrm{d}T'} \  \mathrm{d}T' + \int_{T_{LHB}}^T \frac{\mathrm{d}N_{m>}^*}{\mathrm{d}T'} \  \mathrm{d}T' \nonumber \\
= & \ C_2 + f_{LHB} \ f_{cd} \ (T-T_{LHB}) \ + \nonumber \\
& \ \frac{C_{pd} \ \tau_{pd}}{\beta_{pd}} \ \Gamma{\left(\frac{1}{\beta_{pd}} , \left(\frac{T_o-T}{\tau_{pd}}\right)^{\beta_{pd}}\right)} - C_3
\label{eq_mcumgn_new}
\end{align}

\noindent where $\Gamma$ is the upper incomplete Gamma function, defined as

\begin{equation}
\Gamma(s,x) = \int_s^{\infty} x^{s-1} \ e^{-x} \mathrm{d}x
\end{equation}

\noindent $C_1$ and $C_2$ remain the same as in Eqns.~\ref{eq_c1} and \ref{eq_c2}, and Eq.~\ref{eq_c3} becomes

\begin{equation}
C_3 = \frac{C_{pd} \ \tau_{pd}}{\beta_{pd}} \ \Gamma{\left(\frac{1}{\beta_{pd}} , \left(\frac{T_o-T_{LHB}}{\tau_{pd}}\right)^{\beta_{pd}}\right)}
\label{eq_c3_new}
\end{equation}

\clearpage


\clearpage


\renewcommand{\thetable}{\arabic{table}}

\begin{table}[h!]
\begin{center}
\vspace*{2in}\begin{tabular}{c|c}
\hline
$T_o$ & 4.56 $\mathrm{Ga}$ \\
$T_{LHB}$ & 4.1 $\mathrm{Ga}$ \\
$\lambda_{LHB}$ & 6.93 $\mathrm{Ga^{-1}}$ \\
$f_{LHB}$ & 2 \\
$\tau_{pd}$ & 0.025 $\mathrm{Ga}$ \\
\hline
\end{tabular}
\end{center}
\caption{Nominal values of key parameters in Eqns.~\ref{eq_mcumln} and \ref{eq_mcumgn}}
\label{table_param}
\end{table}

\clearpage


\renewcommand{\thefigure}{\arabic{figure}}

\begin{figure}[h!]
\centering
\includegraphics[width=5.0in]{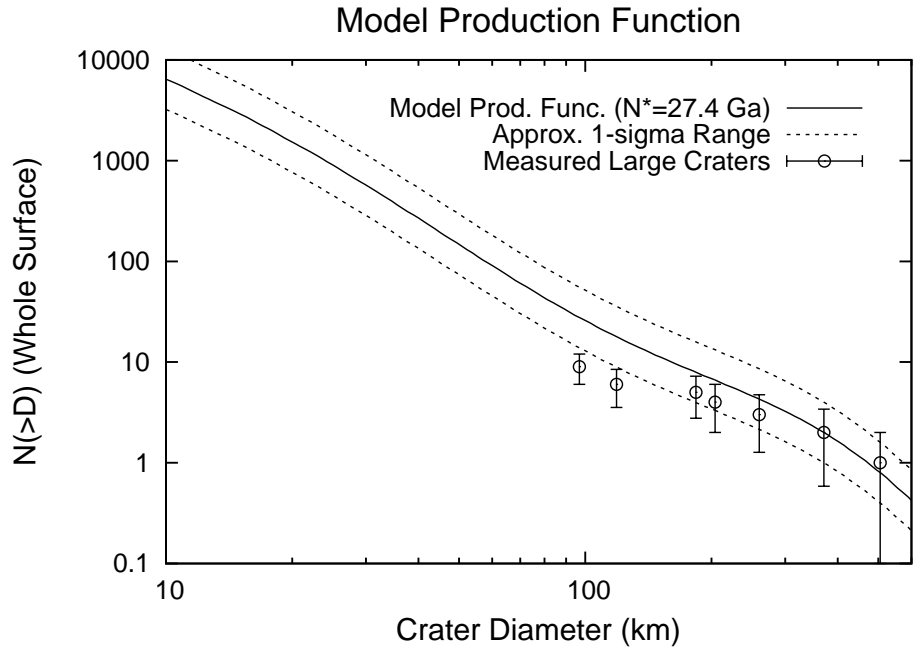}
\caption{The model crater production function $F_{m}(D)$ for Vesta plotted along with the large crater counts of \citet{Marchi2012Sci}.  Error bars for the crater counts are $\sqrt N$.  The solid curve is fit to the two largest craters on Vesta (note that while craters down to $\sim$100 km diameter are shown, they are not included in the fit).  The dashed curves show an approximately 1-sigma range, where an expected value of 1 or 4 large craters is assumed, instead of just two.}
\label{fig_prod_func_fit}
\end{figure}

\clearpage

\begin{figure}[h!]
\centering
\includegraphics[width=5.0in]{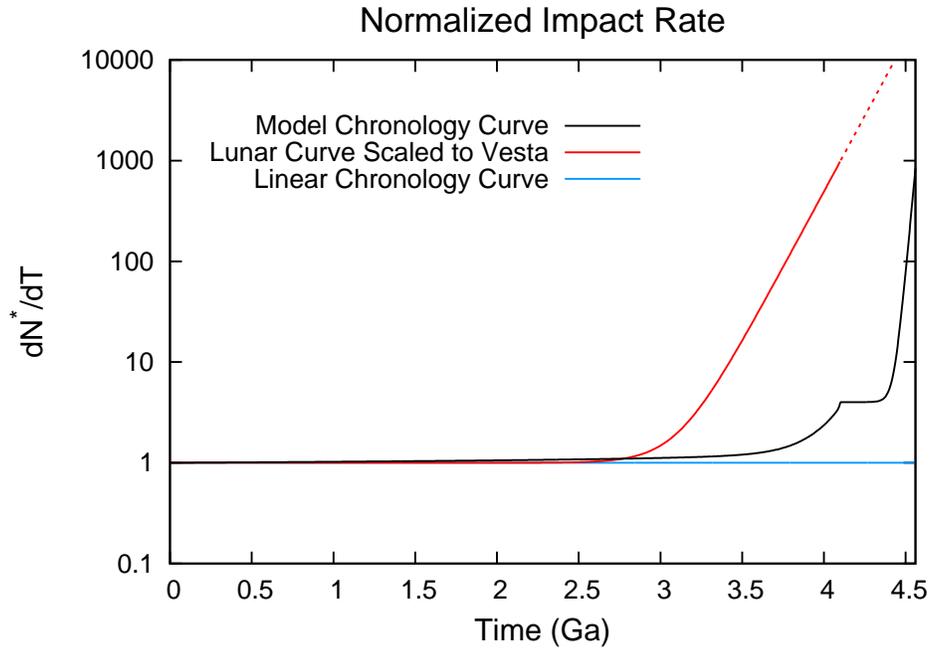}
\caption{The normalized differential production rate of craters $\mathrm{d}N^*/\mathrm{d}T$ in the model chronology (Eqns.~\ref{eq_mdiffnl} and \ref{eq_mdiffng} from Sec.~\ref{sec_m}), for the nominal parameter values given in Sec.~\ref{sec_apply}.  Also shown is the lunar-like chronology (Eq.~\ref{eq_lldiffn} from Sec.~\ref{sec_lun}) and a constant linear production rate curve.  The dashed part of the lunar-like curve prior to 4.1 Ga is an extrapolation, since the lunar chronology on which it is based is only constrained back to $\sim$4.1 Ga.}
\label{fig_diff}
\end{figure}

\clearpage

\thispagestyle{empty}

\begin{figure}[h!]
\centering
\vspace*{-0.5in}\includegraphics[width=4.8in]{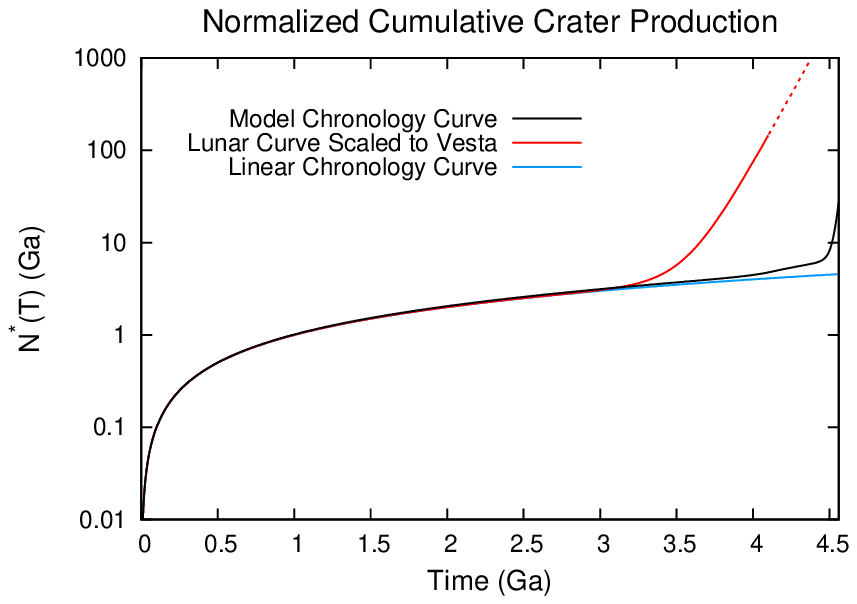}
\includegraphics[width=4.8in]{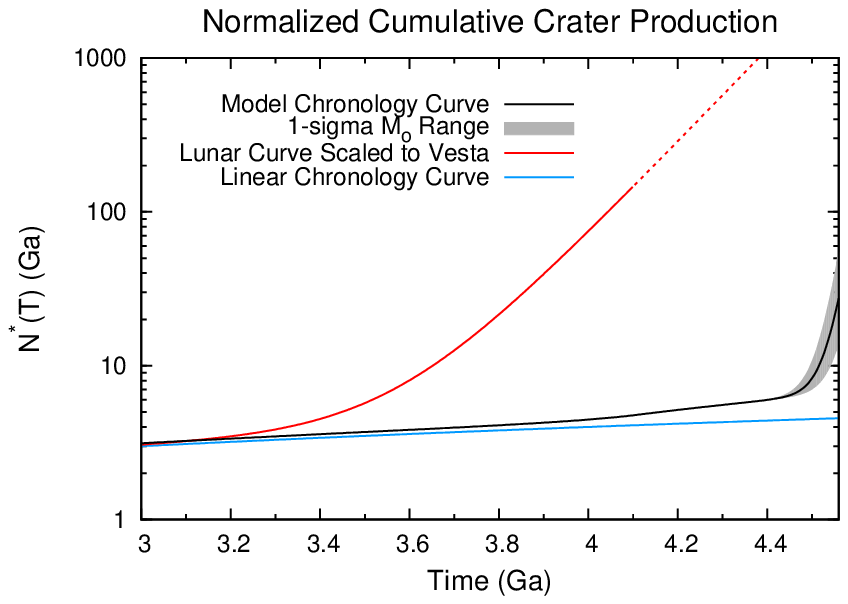}
\caption{The normalized chronology function $N^*(T)$ in the model chronology (Eqns.~\ref{eq_mcumln} and \ref{eq_mcumgn} from Sec.~\ref{sec_m}), for the nominal parameter values given in Sec.~\ref{sec_apply}.  Also shown is the lunar-like chronology (Eq.~\ref{eq_llcumn} from Sec.~\ref{sec_lun}) and a linear chronology curve.  These are the integrals of the differential curves in Fig.~\ref{fig_diff}.  The dashed part of the lunar-like curve prior to 4.1 Ga is an extrapolation, since the lunar chronology on which it is based is only constrained back to $\sim$4.1 Ga.  The bottom figure is the same as the top, but focuses on times greater than 3 Ga, when the curves begin to significantly diverge, and also shows an approximately 1-sigma range based on uncertainties in the initial impacting mass $M_o$.}
\label{fig_cum}
\end{figure}

\clearpage

\thispagestyle{empty}

\begin{figure}[h!]
\centering
\vspace*{-0.5in}\includegraphics[width=4.8in]{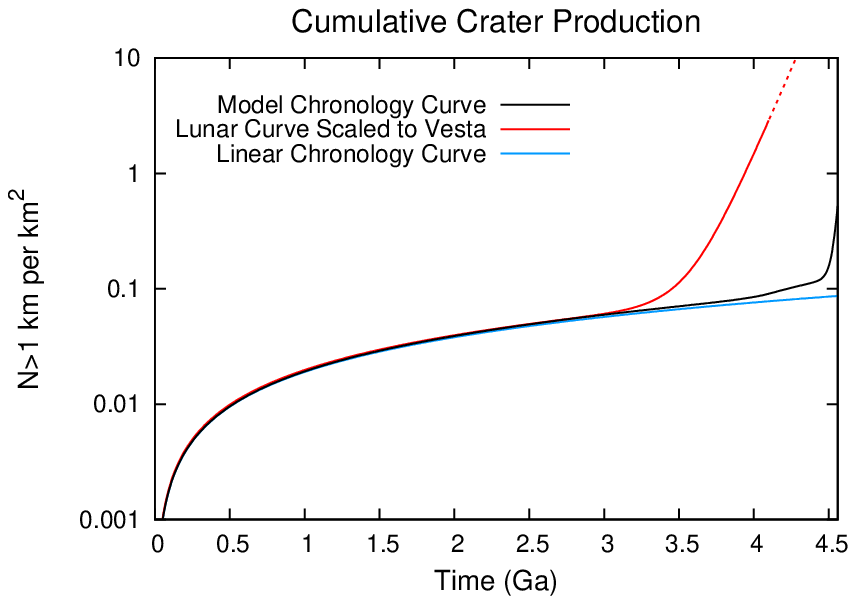}
\includegraphics[width=4.8in]{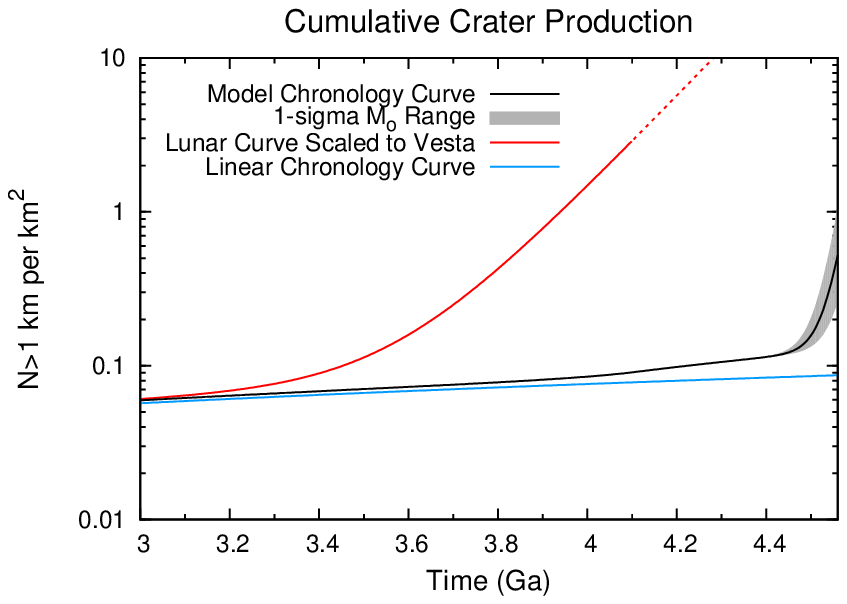}
\caption{Model chronology curve for the density of 1 km and larger craters on Vesta, using the model chronology function from Eqns.~\ref{eq_mcumln} and \ref{eq_mcumgn} (with the nominal parameter values given in Sec.~\ref{sec_apply}) and the model production function $F_{m}(D)$.  Also shown is the lunar-like chronology curve (Eq.~\ref{eq_llcum}) and a linear chronology curve.  The dashed part of the lunar-like curve prior to 4.1 Ga is an extrapolation, since the lunar chronology on which it is based is only constrained back to $\sim$4.1 Ga.  The bottom figure shows the same curves as the top, but focuses on times greater than 3 Ga, when the curves begin to significantly diverge, and also shows an approximately 1-sigma range based on uncertainties in the initial impacting mass $M_o$.}
\label{fig_n1km}
\end{figure}

\clearpage

\begin{figure}[h!]
\centering
\includegraphics[width=4.8in]{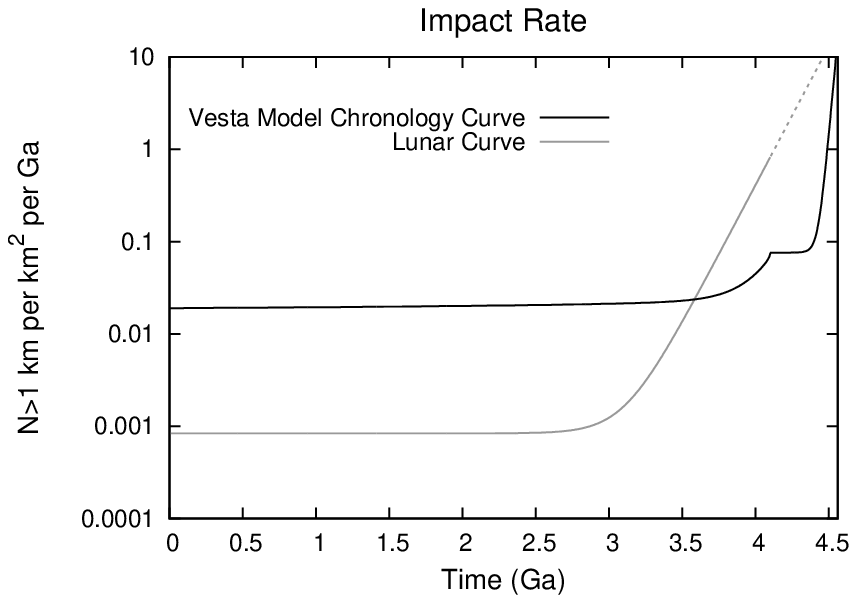}
\includegraphics[width=4.8in]{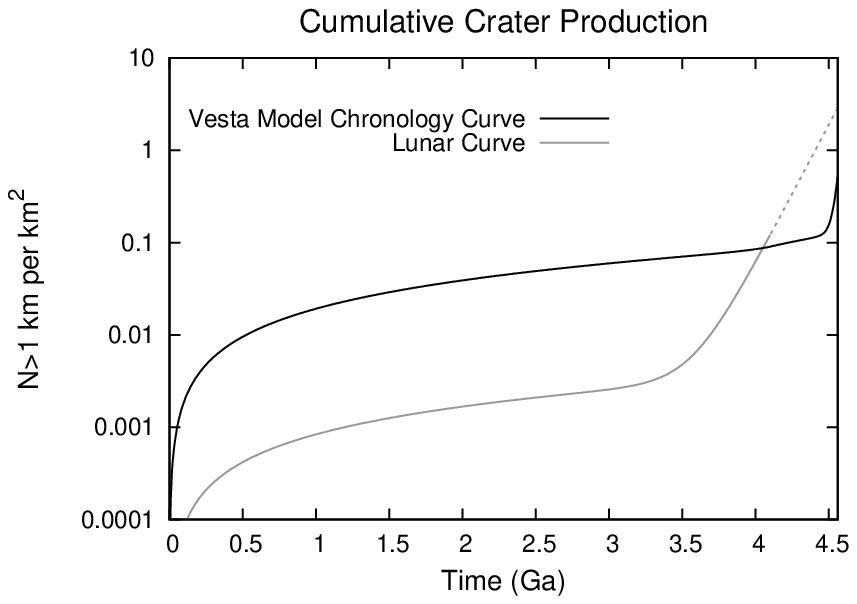}
\caption{Comparison of the model chronology with the lunar chronology.  The top figure shows the formation rate of 1 km and larger craters for Vesta and the moon, while the bottom figure shows the cumulative crater production (ie.~the chronology curve).  The dashed part of the lunar curve prior to 4.1 Ga is an extrapolation, since the lunar chronology is only constrained back to $\sim$4.1 Ga.}
\label{fig_vestluncomp}
\end{figure}
\clearpage

\thispagestyle{empty}

\begin{figure}[h!]
\centering
\vspace*{-0.5in}\includegraphics[width=4.8in]{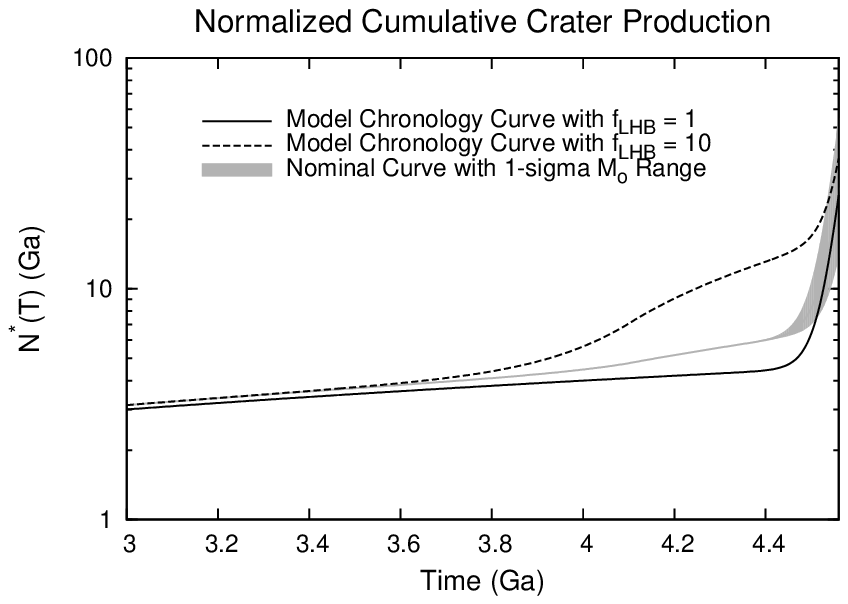}
\includegraphics[width=4.8in]{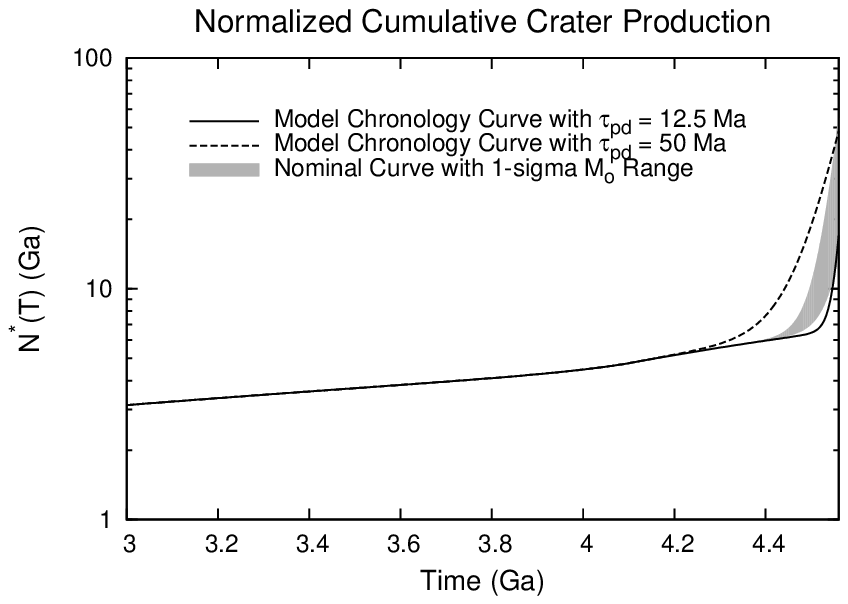}
\caption{Similar to Fig.~\ref{fig_cum}, but showing the effects of varying several of the parameters from the nominal values given in Table \ref{table_param}.  The top figure shows the effect of varying $f_{LHB}$ from 1 to 10, relative to its nominal value of 2.  The case of $f_{LHB}$ = 1 corresponds to no depletion and hence no change in the impact rate at $t_{LHB}$, the only change in the impact rate occurs during the early primordial depletion phase.  $f_{LHB}$ = 10 is consistent with the early Nice Model simulations \citep{Gomes2005Nat}, before subsequent modeling revised that value downwards.  The bottom figure shows the effect of varying the primordial depletion timescale $\tau_{pd}$ from 12.5 Ma to 50 Ma, relative to the nominal value of 25 Ma.  For comparison, the grey curves in both plots show the nominal chronology curve with an approximately 1-sigma range based on uncertainties in the initial impacting mass $M_o$.}
\label{fig_cum_range}
\end{figure}

\clearpage

\begin{figure}[h!]
\centering
\includegraphics[width=5.0in]{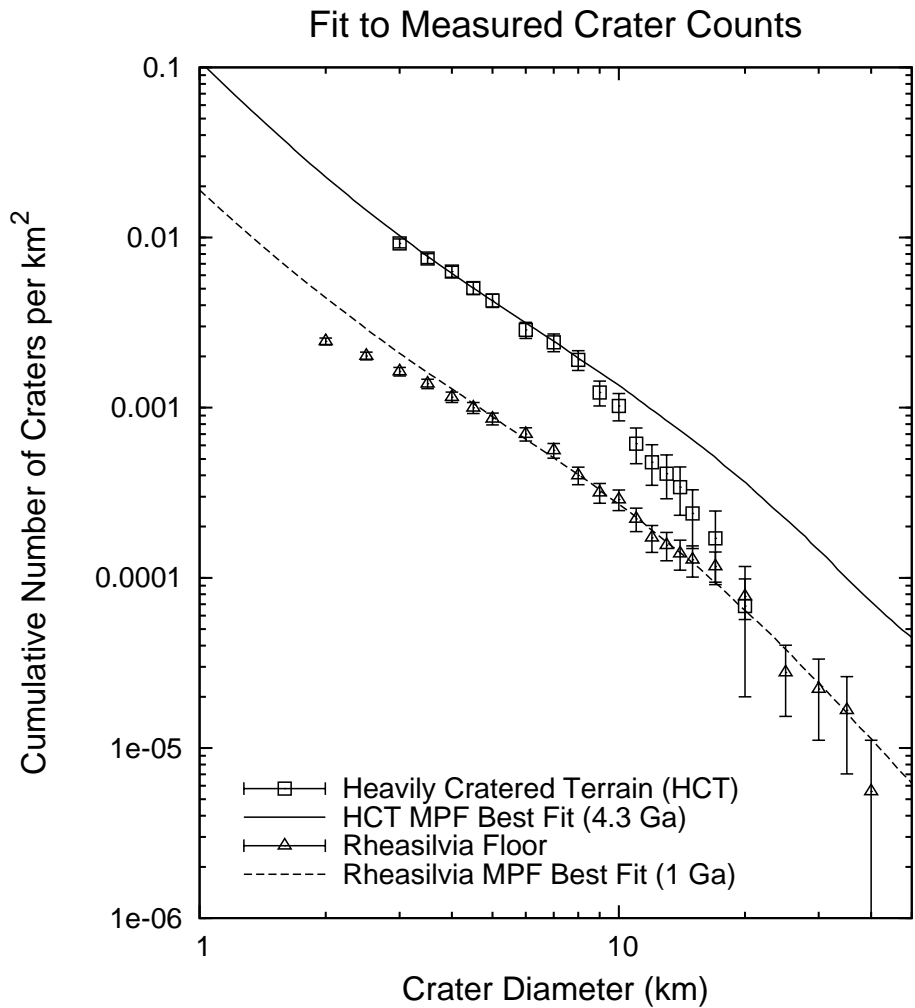}
\caption{Crater counts for two regions on Vesta, with age estimates using the model chronology.  HCT refers to the highly-cratered terrain identified by \citet{Marchi2012Sci}.}
\label{fig_vesta_cc}
\end{figure}

\end{document}